\documentclass[preprint,preprintnumbers,amsmath,amssymb]{revtex4}

\usepackage{graphicx}
\usepackage{dcolumn}
\usepackage{bm}

\begin{document}

\preprint{Proceedings 10th Int. Symp. on Origin of Matter and Evolution of Galaxies OMEG07, Sapporo/ Japan}

\title{Recent Efforts in Data Compilations for Nuclear Astrophysics}

\author{I. Dillmann}\email{iris.dillmann@ph.tum.de}
 \affiliation{Institut f\"ur Kernphysik, Forschungszentrum Karlsruhe, Postfach 3640, D-76021 Karlsruhe, Germany}
 \affiliation{Physik Department E12, Technische Universt\"at M\"unchen, D-85748 Garching, Germany}

\begin{abstract}
Some recent efforts in compiling data for astrophysical purposes are introduced, which were discussed during a JINA-CARINA Collaboration meeting on \textit{"Nuclear Physics Data Compilation for Nucleosynthesis Modeling"} held at the ECT* in Trento/ Italy from May 29th- June 3rd, 2007. The main goal of this collaboration is to develop an updated and unified nuclear reaction database for modeling a wide variety of stellar nucleosynthesis scenarios. Presently a large number of different reaction libraries (REACLIB) are used by the astrophysics community. The "JINA Reac\-lib Database" on \textit{http://www.nscl.msu.edu/\~{}nero/db/} aims to merge and fit the latest experimental stellar cross sections and reaction rate data of various compilations, e.g. NACRE and its extension for Big Bang nucleosynthesis, Caughlan and Fowler, Iliadis et al., and KADoNiS. 

The KADoNiS (Karls\-ruhe Astrophysical Database of Nucleosynthesis in Stars, \textit{http://nuclear-astrophysics.fzk.de/kadonis}) project is an online database for neutron capture cross sections relevant to the $s$ process. The present version v0.2 is already included in a REACLIB file from Basel university (\textit{http://download.nucastro.org/astro/reaclib}). The present status of experimental stellar $(n,\gamma)$ cross sections in KADoNiS is shown. It contains recommended cross sections for 355 isotopes between $^{1}$H and $^{210}$Bi, over 80\% of them deduced from experimental data. 

A "high priority list" for measurements and evaluations for light charged-particle reactions set up by the JINA-CARINA collaboration is presented. The central web access point to submit and evaluate new data is provided by the Oak Ridge group via the \textit{http://www.nucastrodata.org} homepage. "Workflow tools" aim to make the evaluation process transparent and allow users to follow the progress.
\end{abstract}

\maketitle

\section{Stellar neutron capture compilations}

\subsection{Status of stellar $(n,\gamma)$ cross sections \label{exp}}
The pioneering work for stellar neutron capture cross sections was published in 1971 by Allen and co-workers \cite{alle71}. In this paper the role of neutron capture reactions in the
nucleosynthesis of heavy elements was reviewed and a list of recommended (experimental or semi-empirical) Maxwellian averaged cross sections at $kT$= 30 keV (MACS30) presented for nuclei between Carbon and Plutonium.

The idea of an experimental and theoretical stellar neutron cross section database was picked up again by Bao and K\"appeler \cite{bao87} for $s$-process studies. This compilation published in 1987 included cross sections for ($n,\gamma$) reactions (between $^{12}$C and $^{209}$Bi), some ($n,p$) and ($n,\alpha$) reactions (for $^{33}$S to $^{59}$Ni), and also ($n,\gamma$) and ($n,f$) reactions for long-lived actinides. A follow-up compilation was published by Beer et al. in 1992 \cite{BVW92}.

In the update of 2000 the Bao compilation \cite{bao00} was extended down to $^{1}$H and -- like the original Allen paper -- semi-empirical re\-commended values for nuclides without experimental cross section information were added. These estimated values are normalized cross
sections derived with the Hauser-Feshbach code NON-SMOKER \cite{rath00}, which account for known systematic deficiencies in the nuclear input of the calculation. Additionally, the database provided stellar enhancement factors and energy-dependent MACS for energies between $kT$= 5 keV and 100 keV.

The \textsc{KADoNiS} (Karls\-ruhe Astrophysical Database of Nucleosynthesis in Stars) project \cite{kado06} is based on these previous compilations and aims to be a regularly updated database. The current version \textsc{KADoNiS} v0.2 (January 2007) is already the second update and includes -- compared to the previous Bao et al. compilation \cite{bao00} -- 38 updated and 14 new recommended cross sections. A paper version of \textsc{KADoNiS} ("v1.0") is planned for 2008, which also will -- like the first Bao compilation from 1987 \cite{bao87} -- include ($n,p$) and ($n,\alpha$) reactions for light isotopes and ($n,\gamma$) and ($n,f$) reactions for long-lived actinides at $kT$= 30~keV. 

In total, data sets are available for 355 isotopes, including 76  radioactive nuclei (22\%) on or close to the $s$-process path. For 12 of  these radioactive nuclei experimental data is available: $^{14}$C, $^{93}$Zr, $^{99}$Tc, $^{107}$Pd, $^{129}$I, $^{135}$Cs, $^{147}$Pm, $^{151}$Sm, $^{154}$Eu, $^{163}$Ho, $^{182}$Hf, and $^{185}$W. The remaining 64 radioactive nuclei are not yet measured in the stellar energy range and are represented only by semi-empirical cross section estimates with typical uncertainties of 25 to 30\%. Almost all ($n, \gamma$) cross sections of the 277 stable isotopes have been measured. The few exceptions are $^{17}$O, $^{36,38}$Ar, $^{40}$K, $^{50}$V, $^{72,73}$Ge, $^{77}$Se, $^{98,99}$Ru, $^{131}$Xe, $^{138}$La, $^{158}$Dy and $^{195}$Pt. Most of these cross sections are difficult to determine because 
they are not accessible by activation measurements or samples are not available in sufficient amounts and/or enrichment for time-of-flight measurements.

\begin{figure}[!htb]
  \includegraphics{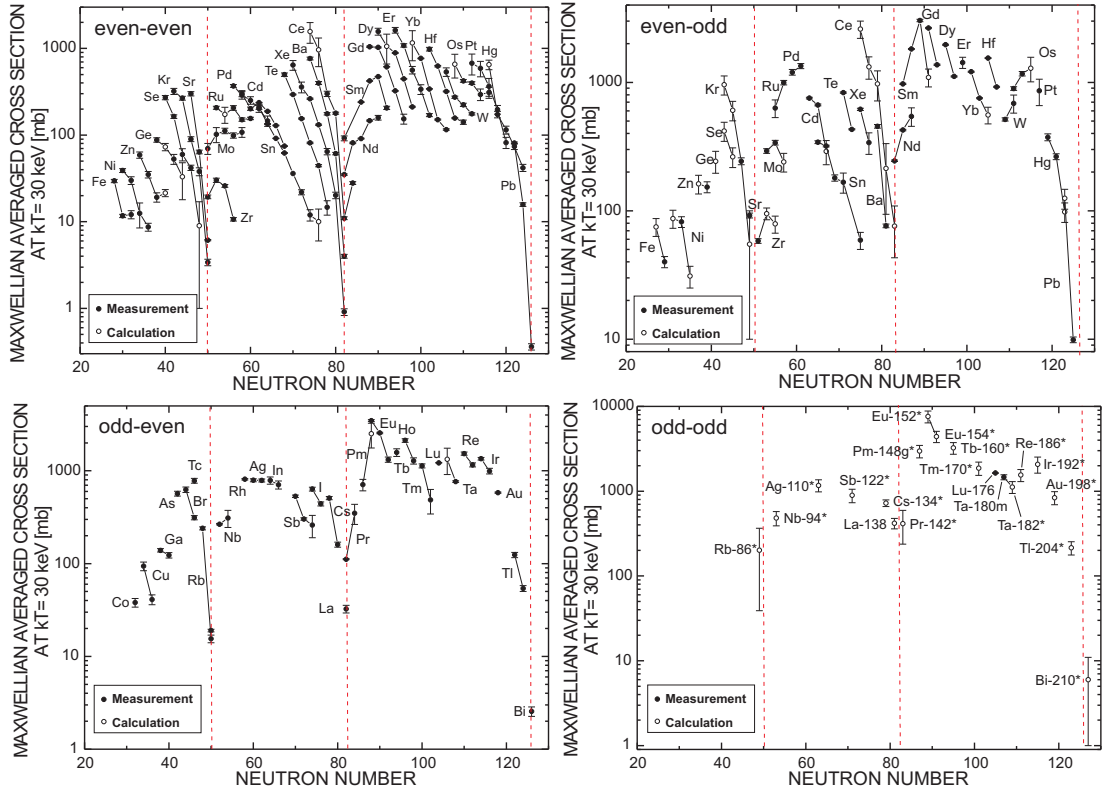}
  \caption{\label{MACS} Maxwellian averaged cross sections at $kT$=30~keV for even-even, even-odd, odd-even, and odd-odd isotopes beyond Iron ($Z$$\geq$26) listed in \textsc{KADoNiS}. Filled circles show measured cross sections, open circles indicate semi-empirical estimates.}
\end{figure}

Fig.~\ref{MACS} shows the status of KADoNiS v0.2 for the mass region above iron ($Z$$\geq$26) at $kT$=30~keV. The even-$N$ nuclei are now almost completely measured, but partially not yet included in the latest KADoNiS version from January 2007. The odd-$N$ nuclei exhibit predominantly only semi-empirical cross sections because most of them are radioactive.

\begin{figure}[!htb]
  \includegraphics{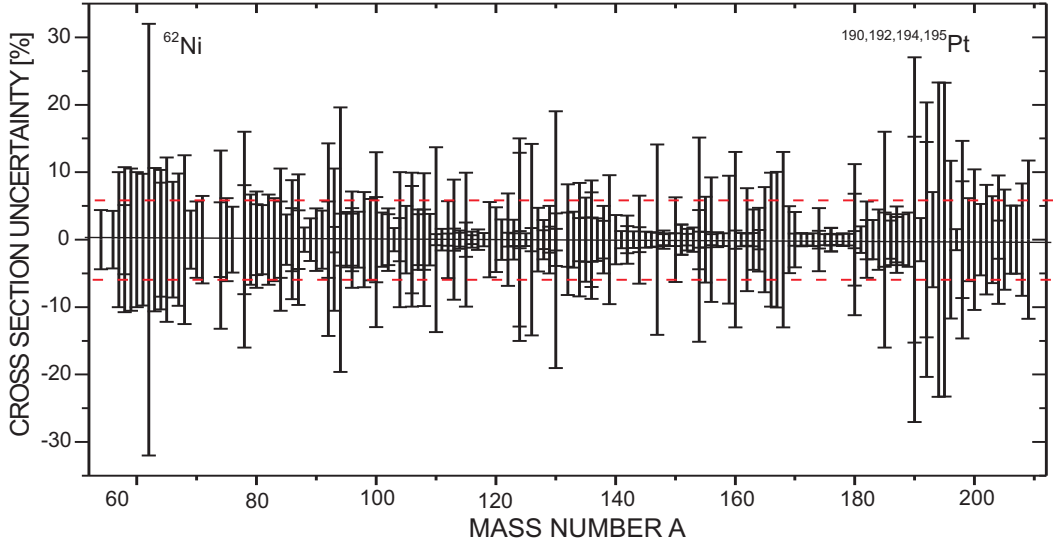}
  \caption{\label{xs-unc} Present experimental uncertainties of stellar ($n,\gamma$) cross sections beyond iron ($Z$$\geq$26) in \textsc{KADoNiS} v0.2 (January 2007). The average uncertainty in this region is $\pm$6\% (without semi-empirical estimates). $^{62}$Ni and some light Pt isotopes exhibit much larger uncertainties.}
\end{figure}

Fig.~\ref{xs-unc} gives the uncertainties at $kT$=30~keV of all experimental $(n,\gamma)$ cross sections measured beyond iron. The average uncertainty in this region is $\pm$6\% (indicated by the vertical dashed lines), but may be considerably larger at lower and higher temperatures. Mavericks with uncertainties $\geq$20\% belong to $^{62}$Ni (where two recent, but contradictory measurements were not yet included \cite{NPA05,TTS05}) and the light Pt isotopes $^{190,192,194,195}$Pt.

\subsection{Extensions of \textsc{KADoNiS}}
Several extensions are planned for \textsc{KADoNiS} in 2008. The $s$-process library will be complemented in the near future by some ($n,p$) and ($n,\alpha$) cross sections measured at $kT$=30 keV, as it was already done in \cite{bao87}. Additionally some "gaps" in the list of stable isotopes will be filled. For different reasons in the previous compilations the datasets for $^{2}$H, $^{6}$Li, $^{9}$Be, $^{10,11}$B, $^{17}$O, and $^{138}$La were missing. 

Furthermore it is planned to include more radioactive isotopes, which are relevant for $s$-process nucleosynthesis at higher neutron densities (up to 10$^{11}$ cm$^{-3}$). These isotopes are more than one unit away from the "classical" $s$-process path on the neutron-rich side of stability, and their stellar ($n,\gamma$) cross sections have to be extrapolated from known cross sections with the statistical Hauser-Feshbach model. The present list 
covers 73 new isotopes and is available on the \textsc{KADoNiS} homepage at  
\textit{nuclear-astrophysics.fzk.de/kadonis/new\_iso\_list.txt}. However, only a few of these radioactive isotopes can be measured with present techniques (e.g. $^{60}$Fe), the largest fraction has to wait for future radioactive ion beam (RIB) facilities.

Once all available datasets are implemented in the $s$-process database, a re-calculation of semi-empirical estimates based on the latest experimental results of neighboring nuclides will be performed and \textsc{KADoNiS} v1.0 can be published. 

The second big part of the \textsc{KADoNiS} project is an experimental $p$-process database. A test version is already online at \textit{nuclear-astrophysics.fzk.de/kadonis/pprocess/}. This $p$-process database aims to be a collection of the available experimental data close to or within the Gamow window of the $p$ process ($T_9$= 2-3 GK, corresponding to E$_p$$\approx$ 1.5-6 MeV or E$_\alpha$$\approx$ 4-14~MeV). However, the largest fraction of $p$-process reactions concerns short-lived radioactive nuclei, which have to be inferred from theoretical work, e.g. the Hauser-Feshbach statistical model \cite{hafe52,rath00}.

Some experimental information is available for charged-particle reactions, but the largest amount of data concerns $(n,\gamma)$ data, which is connected via detailed balance with the respective $(\gamma,n)$ reactions needed for the "$\gamma$ process" mechanism of the $p$ process. Most of this neutron capture data was measured with the activation technique at one single energy ($kT$=25~keV), and thus has to be extrapolated to the respective $\gamma$-process energies ($kT$=170-260~keV) with the help of energy-dependencies from the Hauser-Feshbach theory.

Of utmost importance in this respect are photodisintegration and neutron-induced particle-exchange reactions. $(\gamma,n)$ and $(n,\gamma)$ reactions influence the reaction flow strongly in the whole $p$-process mass region between $A$=70-208. On the other hand $(\gamma,p)$ and $(n,p)$ are only important for the production of light $p$-process isotopes up to the $N$=82 shell ($^{144}$Sm), whereas $(\gamma,\alpha)$ and $(n,\alpha)$ exhibit a very strong influence only for heavy $p$ isotopes \cite{RGW06,ID06}.  

\section{Other JINA-CARINA collaboration efforts}
\subsection{Light charged-particle reactions}
The JINA-CARINA collaboration has set up a priority list for measurements and evaluations of light charged-particle reactions. Several reactions were grouped together according to their priority, and the evaluation and experimental status was discussed. The 20 reactions classified with "high priority" are listed in Table~\ref{tab:prio}. Among them are the 3$\alpha$ and $^{12}$C$(\alpha,\gamma)$ reactions, as well as the neutron source reactions for the $s$ process, $^{13}$C$(\alpha,n)$ and $^{22}$Ne$(\alpha,n)$.

\begin{table}[!htb]
\begin{tabular}{cccc}
\hline
  \multicolumn{4}{c}{Reaction} \\
  $(p,\gamma)$ & $(p,\alpha)$ & $(\alpha,x)$ & $(\alpha,n)$  \\
\hline
$^{7}$Be$(p,\gamma)$$^{8}$B  & $^{17}$O$(p,\alpha)$$^{14}$N & D$(\alpha,\gamma)$$^{6}$Li & $\alpha$$(\alpha,n)$$^{7}$Be   \\
$^{14}$N$(p,\gamma)$$^{15}$O & $^{18}$F$(p,\alpha)$$^{15}$O & 3$\alpha$  & $^{15}$O$(\alpha,n)$$^{18}$Ne \\
$^{21}$Na$(p,\gamma)$$^{22}$Mg&& $^{12}$C$(\alpha,\gamma)$$^{16}$O & $^{17}$O$(\alpha,n)$$^{20}$Ne \\
$^{24}$Mg$(p,\gamma)$$^{25}$Al & &$^{14}$O$(\alpha,p)$$^{17}$F & $^{13}$C$(\alpha,n)$$^{16}$O\\
$^{25}$Mg$(p,\gamma)$$^{26}$Al & & $^{18}$Ne$(\alpha,p)$$^{21}$Na & $^{22}$Ne$(\alpha,n)$$^{25}$Mg \\
$^{25}$Al$(p,\gamma)$$^{26}$Si & & & $^{25}$Mg$(\alpha,n)$$^{28}$Si \\
$^{26}$Al$(p,\gamma)$$^{27}$Si & & & \\
\hline
\end{tabular}
\caption{"High priority" list for measurements and evaluations of light charged-particle reactions published by the JINA-CARINA Collaboration.}
\label{tab:prio}
\end{table}

\subsection{Unified reaction rate library}
The main problem of nucleosynthesis modeling in astrophysics is the diversity of available reaction rate libraries (REACLIBs) depending on the respective requirements. More or less every group uses its "own" library, which makes inter-comparisons of different modeling results impossible. The most widely used libraries are based on SMOKER or NON-SMOKER predictions \cite{rath00} with some experimental information of light isotopes (\textit{http://download.nucastro.org/astro/reaclib}). Additionally the reaction libraries include (partially) experimental results or fitting parameters from NACRE \cite{NACRE} (\textit{http://pntpm.ulb.ac.be/Nacre/ nacre\_d.htm}) and its extension for Big Bang nucleosynthesis \cite{BBN} (\textit{http://pntpm.ulb.ac.be/bigbang}), Caughlan and Fowler \cite{CF88}, Iliadis et al. \cite{IDS01}, or very recently KADoNiS \cite{kado06} (\textit{http://nuclear-astrophysics.fzk.de/kadonis}). 

The aim of the collaboration is to merge all of these different libraries and include them in the unified JINA Reaclib Database (\textit{http://www.nscl.msu.edu/\~{}nero/db/}). In this context the latest available experimental results (which also have been evaluated) will be fitted and provided in the most recent REACLIB version, but also previous libraries can be downloaded from this website.

A reaction rate library from Basel university (\textit{http://download.nucastro.org/astro /reaclib}) has been recently updated with cross sections from \textsc{KADoNiS} v0.2 \cite{ID06}. In a first approach the existing fitting parameters (based on NON-SMOKER cross sections \cite{rath00}) for the $(n,\gamma)$ and $(\gamma,n)$ rates were only normalized to the \textsc{KADoNiS} cross sections at $kT$=30~keV. These 355 isotopes have to be checked again for cross sections measured with the time-of-flight technique, which exhibit a proper energy-dependence. In these cases a careful refitting is necessary, and the new fit parameters will be added into the respective updated REACLIB.

\subsection{Workflow tools}
The central access point for submission and evaluation of new experimental data is provided via \textit{http://nucastrodata.org/infrastructure.html}. By launching the interface, the display for "Computational Infrastructure for Nuclear Astrophysics" is opened, where users can browse the status of evaluations. On the same website "Scientific contributors" can send their data to evaluators/ referees/ editors, which will comment on this. Flowcharts will show the progress to ensure the transparency of the whole evaluation process.

\section{Summary}
Presently a lot of effort is ongoing concerning a transparent evaluation progress of reaction rates and the allocation of a unified reaction library for nucleosynthesis modeling. Central web access points provided by the JINA-CARINA collaboration are \textit{http://nucastrodata.org/} and \textit{http://www.nscl.msu.edu/\~{}nero/db/}. A priority list for measurements and evaluations of light charged-particle reactions has been set-up by the collaboration.

\begin{acknowledgments}
I.D. was supported by the Swiss National Science Foundation Grants 2024-067428.01 and 2000-105328.
The Joint Institute for Nuclear Astrophysics (JINA) is supported by the National Science Foundation through the Physics Frontier Center Program.
The CARINA network (Challenges and Advanced Research In Nuclear Astrophysics) is supported by the EURONS through the 6th EC Research Framework Program (FP6).
\end{acknowledgments}

\end{document}